# Time Parallel Scalable Multiphysics/Multiscale Framework

George Frantziskonis, Krishna Muralidharan and Pierre Deymier

University of Arizona

Srdjan Simunovic and Sreekanth Pannala

Oak Ridge National Laboratory

We propose a new computational framework that combines the recently developed time-parallel (TP) and the compound wavelet matrix (CWM) methods. The framework, termed tpCWM, offers significant computational acceleration by making multiscale/multiphysics simulations computationally scalable in time and space domains. We demonstrate the accuracy and the scalability of the method on a prototype problem with oscillatory trajectory. The method corrects the coarse solution by iterative use of the CWM, which compounds the fine and the coarse solutions for the processes. Computational savings, over the fine solution as well as the TP method, in terms of the real time required to perform the simulations, can reach several orders of magnitude. We believe that this method is general enough to be applicable to a wide-class of computational physics problems. Tendency towards large number of cores and processors in parallel computers is compatible with the computational scalability of the algorithm.

The most challenging problems in computational science involve phenomena coupled over several orders of magnitude in temporal and spatial scales. Domain decomposition and multigrid methods address primarily the spatial aspect of the computational acceleration, while temporal acceleration, if feasible, is usually governed by the smallest time scales in the problem. The latter is a serious computational hurdle, compounded by their direct link to the spatial resolution.

For multiphysics problems temporal scaling is important due to different operating physical processes evolving and interacting at different time and spatial scales. Current time acceleration schemes, such as Multiple Time Stepping (MTS) and Spectral Deferred Correction (SDC) methods[1], still treat time in an incremental manner. These methods are primarily applicable for formulations with scale separation.

Time parallel (TP) algorithms [2-4] treat time in a similar manner as domain decomposition algorithms. A multiphysics/multiscale framework capable of temporal and spatial scaling, namely the compound wavelet matrix (CWM) [3, 5-7] has been developed. We propose the combination of TP and CWM methods, termed tpCWM, which constitutes a multiscale framework that is computationally scalable in both space and time.

The key idea in TP algorithms is to use different time propagators distributed across phase space, and iterate on their evolution until convergence. In the case of two propagators, termed here "coarse" and "fine," in each global iteration one obtains a temporally coarse solution of a problem, and then at several temporal "nodes" along the coarse solution instantiate fine-grained temporal simulations. The fine simulations correct the coarse one, and the process is repeated, in a predictor-corrector sense, until convergence is achieved. The method is easily parallelizable and conceptually simple, clear advantages for utilizing supercomputing resources. The solution at the *k+1* TP iteration and *n+1* temporal node is given by:

$$y_{n+1}^{k+1} = C(y_n^{k+1}) + \left[ F(y_n^k) - C(y_n^k) \right], \tag{1}$$

where, *C* and *F* stand for coarse and fine projectors respectively. For this procedure to be effective, the coarse projector has to be computationally cheap as it constitutes the serial part at each TP iteration while the fine propagator is performed in parallel on each of the processors. For tpCWM, where (1) becomes $y_{n+1}^{k+1} = C(y_n^{k+1}) + \left[ CWM(y_n^k) - C(y_n^k) \right]$, the convergence proofs for TP hold, for the case of clearly separated scales with quasi-stationary fine processes. For a general case, we do not have a mathematical proof at this point and we resort to computational experiments such as those reported in this paper. Figure 1 shows a schematic of the TP and tpCWM solution process; the fine solutions are iterated in coordination with the coarse one, in a predictor corrector sense, until satisfactory convergence occurs.

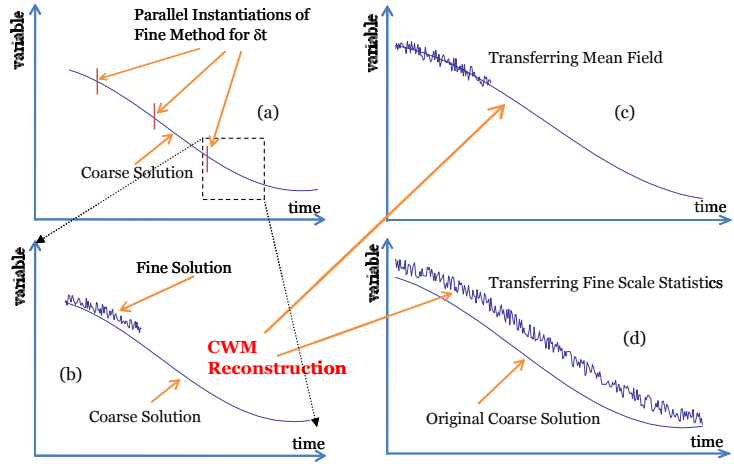

**FIG 1 Schematic of the TP and CWM methods. (a) The TP method. The fine method instantiates at several temporal "nodes" typically for a period δt that covers**

**time until the next node. (b) The temporal CWM. The fine method is employed for a fraction of the coarse method for each of the temporal nodes. (c) The CWM reconstruction updates the mean field. (d) The CWM reconstruction updates the temporal fluctuations.**

The idea behind CWM method [5-8] is to first obtain the solution of a problem at multiple scales, and fuse the results into a global-compound-matrix in the wavelet domain. The CWM allows for combining multiphysics and multiscaling in a single framework across the entire temporal and spatial domains of the problem.

Figure 1 shows the CWM operating on temporal scales. By compounding small-scale information from the fine solution and large-scale information from the coarse method, the improved temporal response is obtained that not only "corrects" the coarse trajectory, but also incorporates small-scale information from the fine method.

We consider a multiscale chemical reaction problem in which the coarse propagator is a solution of a set of deterministic, ordinary differential equations (e.g. rate equations), and the fine propagator is a solution of stochastic method (e.g. kinetic Monte Carlo, KMC)[9]. The benchmark solution is the fine propagator (KMC), ran over the entire time interval.

Let a, b denote two time-dependent concentrations of the two reactive species. At steady-state, the concentrations are $a_0, b_0$, and deviations from steady state are denoted as $A = a - a_0$, $B = b - b_0$, respectively. Let us consider the reaction rate ODE equations of the following form:

$$\frac{dA}{dt} = \kappa_{11}A + \kappa_{12}B \quad , \quad \frac{dB}{dt} = \kappa_{21}A + \kappa_{22}B \qquad (2)$$

Analytical solution of (2) for $\kappa_{11} = \kappa_{22} = 0$, $-\kappa_{21} = \kappa_{12} = \kappa = 0.001 s^{-1}$, and initial values $A_0=0$ and $B_0=10000$, yields oscillatory solutions for A, and B, as $A(t) = B_0 sin(\kappa t)$.[10]

The coarse model uses a deterministic algorithm for solving the ODE system (2). The first-order Euler scheme yields, with $\Delta$ denoting finite difference

$$\Delta A = \kappa B \Delta t, \qquad \Delta B = -\kappa A \Delta t \qquad (3)$$

Large time increments are used as a prototype coarse method with large, yet stable, error in order to examine the tpCWM method. In fact, the solution diverges with time, and yet, despite the divergence, it is shown that the tpCWM converges to the correct solution very quickly.

The KMC algorithm is the fine propagator for the kinetic evolution (2) for the deviations from the steady state. Times required for one unit change in the value of A, B for the oscillatory case are expressed as:

$$t_1 = -\frac{1}{\kappa|A|}\ln(1-R_1) \quad , \quad t_2 = -\frac{1}{\kappa|B|}\ln(1-R_2) \qquad (4)$$

where, $R_1$ and $R_2$ are independent uniformly distributed random numbers between zero and unity. At every KMC iteration step, the minimum of $t_1, t_2$ is the time increment associated with the selected unit change event. We will use the KMC solution over the entire interval as the benchmark.

A TP solution of form (1) applied to (2) calls for instantiation of KMC solutions at the beginning of each of the time increments of the coarse method, called nodes. The number of nodes is considered to be, for the present case equal to the number of TP processes (number of processors) denoted as $n_p$. Each KMC simulation is performed for the time interval between the nodes for every TP iterate. At the end of each iteration, the error of each KMC run (the second term on the RHS of Eq. 1) is evaluated as the difference between the KMC solution and the coarse solution at the same iteration level, which then is used at each time increment to correct the coarse solution at the new TP iterate. The interaction of the fine and coarse method during the iteration process allows for very fast convergence of the method.

A tpCWM solution to (2) calls for instantiation of CWM solutions at each node. For the first iteration, the initial conditions needed in the CWM are obtained from the coarse solution. For subsequent iterations, the tpCWM algorithm is identical to the TP one described above.

There are some important differences between the TP and the tpCWM schemes. First, the CWM for each node is such that the KMC is only run for a fraction of the interval between the time nodes. This fraction, denoted as $f$ was herein chosen to be 1/16.

Figure 2 shows results from the tpCWM process for 3 iterations, where, similarly to the TP, it can be seen that it only takes a few iterations for this problem to converge.

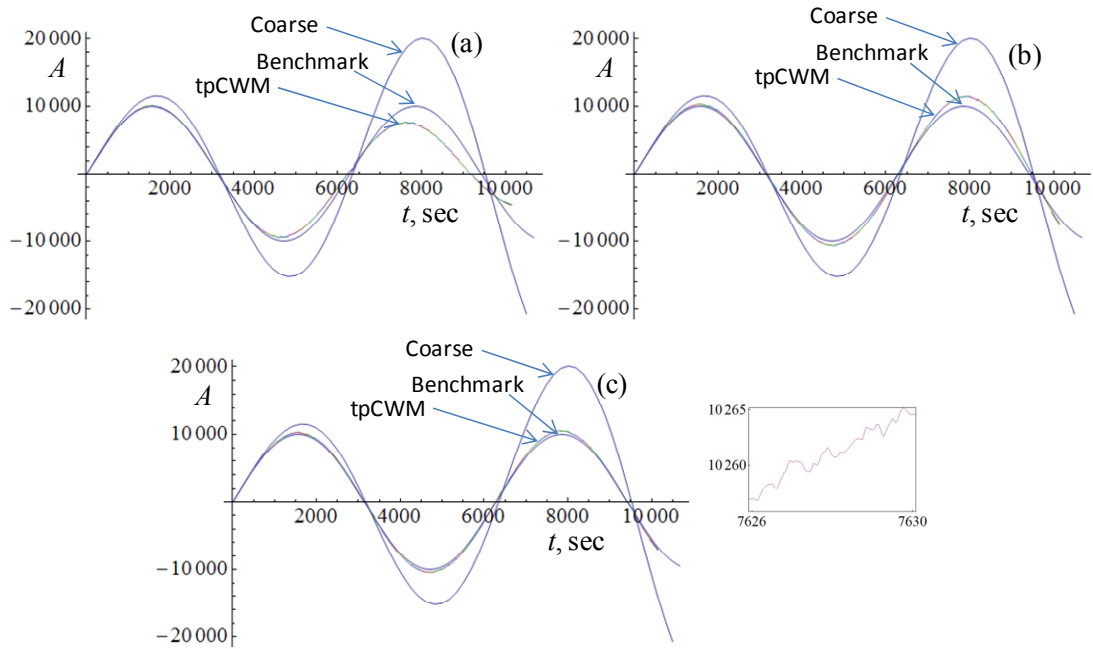

**FIG 2 tpCWM solution, $n_p$=60, at iterations 1 (a), 2 (b), 3 (c). The CWM, for a particular time interval is shown in (c) (inset) depicting the relevant fluctuations. Also the benchmark solution is shown, towards which the tpCWM iterations converge.**

Figure 3a shows the relative error (measured by the L-2 norm normalized with respect to the error at the first iteration) of the concentration of species $A$, with the number of iterations. Solution converges to the stochastic noise floor (due to KMC) after 3-4 iterations, for a reasonable number of processors $n_p$ and a ratio $f$ that is not too small.

Figure 3b shows the relative error for $n_p$=30, $f$=1/16, and time increment in the coarse method of 350 s, showing the rapid convergence for even such a "crude" coarse solution and small number of processors. For the long-time behavior (t=15000 s), the coarse solution diverges with an error of over 250% whereas the tpCWM converges in 3 iterations.

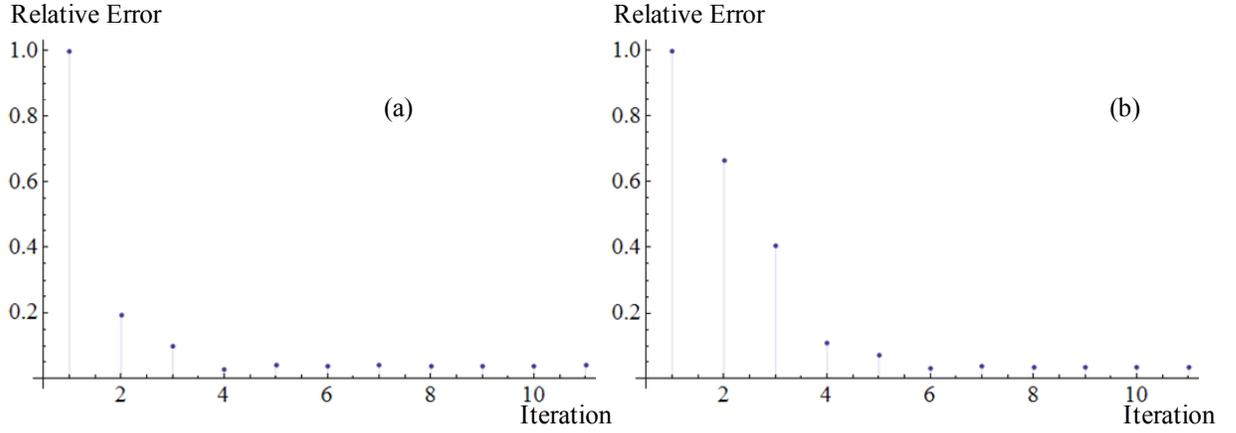

**FIG 3 Relative error as a function of the number of iterations for (a) using 60 TP processes, i.e. $n_p = 60$; (b) $n_p = 30$.**

The computational advantage of tpCWM stems from the efficiency of the CWM propagator as it needs only a fraction of the time interval between the nodes to be solved by KMC, versus a full interval run of KMC in the TP method. Thus, it adds to the computational acceleration from the use of parallel processors at every iteration. For example, for $f=1/16$ and 7 iterations, the computational savings of the tpCWM over TP is approximately 7*16~112 times. We have also found that tpCWM requires fewer global iterations than TP method for the same level of convergence due to the ability of CWM to project the mean tendency of the signal. The cost of the CWM wavelet transform calculations reduces this factor, and in the above example making the actual saving factor ~95 instead of 112.

Let $n_i$ denote the number of iterations required for convergence, considered here to be reached when the relative error is 0.01. The ratio $r$ of TP processes over $n_i$, denoted as

$$r = \frac{n_p}{n_i} \qquad (5)$$

is 60/3 for this particular application, for $n_p$=60, of the tpCWM method presented above. Figure 4, shows the factor of computational savings *X*, defined as the ratio of computational time required for the benchmark method (pure KMC) over time required for the tpCWM, as a function of *r* (number of processors/number of iterations) and *f* (fraction of KMC time used in each assigned time interval). Three orders of magnitude in *X* can be achieved by *r* in the range of 20 and *f* in the order of 1/64. The role of *r* is important, as its increase with the number of processes will indicate computational scalability of the method. Based on results from using 30, 60, and 90 TP processes $n_p$, the required number of iterations was tracked, and they were 6, 4, and 3 respectively, showing that the number of iterations remains low and even becomes lower as $n_p$ increases.

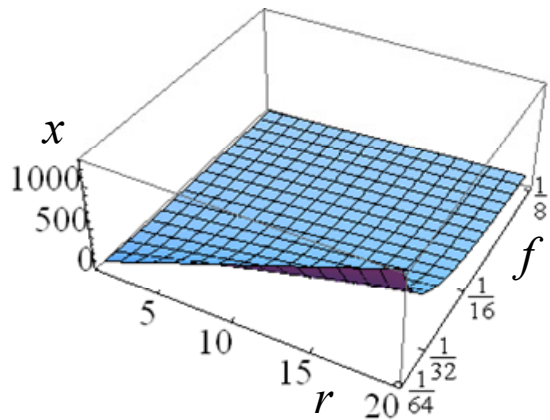

**FIG 4 Factor of computational savings, *X*, as a function of the ratio *r* and the fraction *f*.**

Our analysis indicates that the combination of TP and CWM, termed herein as tpCWM, enables significant computational acceleration for multiscale problems. Major advantages of tpCWM over the TP method are the realization of computational savings at every iteration step, and the computational scalability with the increasing number of processors. The CWM corrects the coarse and fine solutions before they are used in each of the TP steps, and provides the TP efficient interaction of the fine and coarse methods over the entire spatial and temporal domains instead of just at their common temporal nodes.